\documentclass[a4paper]{IEEEtran}

\usepackage[hidelinks,bookmarks=false]{hyperref}
\usepackage[official]{eurosym}
\usepackage{flushend,amsmath}

\usepackage[utf8]{inputenc}

\ifCLASSINFOpdf
  \usepackage[pdftex]{graphicx}
  \graphicspath{{graphics/}}
  \DeclareGraphicsExtensions{.pdf,.jpeg,.png}
\else
\fi

\IEEEpubid{978-1-5090-5499-2/17/\$31.00~\copyright~2017 IEEE}

\usepackage[style=ieee]{biblatex}
\bibliography{network_clustering}

\usepackage{tikz}

\usepackage[europeanresistors,americaninductors]{circuitikz}
\usepackage{adjustbox}

\usepackage{booktabs}
\newcommand{\ra}[1]{\renewcommand{\arraystretch}{#1}}

\usepackage{color}

\def\co{CO${}_2$}
\def\el{${}_{el}$}
\def\th{${}_{th}$}

\newcommand{\figscale}{0.8}

\begin{document}
%

\title{The role of spatial scale in joint optimisations of generation
  and transmission for European highly renewable scenarios}

\author{\IEEEauthorblockN{Jonas Hörsch, Tom Brown} \\
  \IEEEauthorblockA{Frankfurt Institute for Advanced Studies, Ruth-Moufang-Straße 1, 60438 Frankfurt am Main, Germany \\
Email: \href{mailto:hoersch@fias.uni-frankfurt.de}{hoersch@fias.uni-frankfurt.de}}
}


%


\maketitle

\begin{abstract}
  The effects of the spatial scale on the results of the optimisation
  of transmission and generation capacity in Europe are quantified
  under a 95\% CO2 reduction compared to 1990 levels, interpolating
  between one-node-per-country solutions and
  many-nodes-per-country. The trade-offs that come with higher spatial
  detail between better exposure of transmission bottlenecks,
  exploitation of sites with good renewable resources (particularly
  wind power) and computational limitations are discussed. It is shown
  that solutions with no grid expansion beyond today's capacities are
  only around 20\% more expensive than with cost-optimal grid
  expansion.
\end{abstract}


%
\IEEEpeerreviewmaketitle

\section{Introduction}

Optimising investment in the electricity system to reduce greenhouse
gas emissions is computationally intensive. Transmission investment
should be jointly optimised with generation investment, so that the
benefits of exploiting the sites with the best renewable resources can
be balanced against the network expansion costs; continental-scale
areas should be considered, so that synoptic-scale weather variations
($\sim$600-1000~km), which particularly affect wind generation, can be
balanced; at the same time, high spatial detail is required to capture
both variations in renewable resources and existing transmission
bottlenecks.

Previous studies have typically sacrificed at least one of these
goals. In some studies, only a single node per country or group of
countries has been considered
\cite{Czisch,Scholz,Rodriguez2013,schlachtberger2017}, ignoring
national transmission networks and local differences in weather
conditions. Other studies consider the transmission network in detail,
but only for single countries \cite{KWK2,Ecofys}, neglecting the
benefits of international cooperation. Other studies maintain both a
pan-continental scope and transmission network detail, but fix the
generation fleet and only optimise the transmission network
\cite{Egerer,Brown}.

In this paper it is attempted to bring a more systematic approach to
the question of spatial resolution in electricity system
optimisations. A clustering methodology called `$k$-means' is used to
successively reduce the number of nodes in the European transmission
network from its full level of spatial detail down to a level where
there is only one node per country. The effects on the results of the
optimal investments in generation and transmission are then studied as
the spatial resolution is changed. A high spatial resolution reveals
transmission bottlenecks that might either restrict welfare-enhancing
transfers or force transmission upgrades; ignoring these effects in a
low resolution model leads to an underestimate of the total costs. In
a low resolution model one is also forced to average the renewable
resources over a larger area, which lowers the average capacity
factors, even with a weighting towards better sites; at high
resolution the sites with the highest capacity factors can be fully
exploited, particularly for wind.

In recent decades as large-scale optimisation has gained in
importance, many methods have been suggested in the literature to
reduce the whole network to a number of clusters, rather than
focussing on a binary exterior-interior division like the Ward or
Radial-Equivalent-Independent methods~\cite{Deckmann1980}. Standard
clustering algorithms from complex network theory \cite{Jain1999} have
been applied on the network structure, including $k$-means clustering
\cite{Hartigan1979} on electrical distance between buses
\cite{TEMRAZ1994301,Blumsack2009,CotillaSanchez} and spectral
partitioning of the Laplacian matrix \cite{Hamon2015}.
Equivalents based on zonal Power Transfer Distribution Factors (PTDFs)
were considered in \cite{Cheng2005}, while a methodology based on
Available Transfer Capacities (ATCs) was developed in
\cite{Shayestah2015}. A more economic focus was taken in
\cite{Singh2005}, where buses were clustered based on similar average
locational marginal price (LMP). The final report of the recent
e-Highways 2050 project \cite{eHighways2050} that considered network
expansion needs in Europe contains both a summary of network
clustering methods and suggestions for mixed metrics combining several
characteristics to define nodal similarity.

\IEEEpubidadjcol

In the following sections we review the clustering methodology, the
investment model, the data input for the European electricity system
and the results for different aggregation scales and levels of grid
expansion.

\section{Methodology for network reduction}\label{sec:reduction}

We first describe the method to derive a clustered equivalent network,
with fewer buses and lines, from a more detailed network. First, the
network buses are partitioned into clusters, then an equivalent
network is constructed with one bus per cluster and aggregated lines
between the new buses.

The network reduction method used here for an equivalent network of
$k$ buses consists of the following steps:

\begin{enumerate}
\item Univalent buses, i.e. network stubs or `dead-ends', are
  aggregated to their neighbours in an iterative process until all
  buses are multi-valent, since such stubs are typically short lines,
  connecting single generators to the main network.
\item The remaining buses labelled by $n$ are assigned a weight $w_n$
  proportional to the load and today's conventional generation
  capacities at the bus and coordinates $x_n$ based on their
  geographical location.
\item The $k$-means algorithm is used to find the geographical
  positions of $k$ centroids $\{ x_c\}$ for $c =1,\dots k$ by
  minimising the weighted sum of squared distances from each centroid
  to its clustered members $N_c$:
  \begin{equation}
    \min_{\{x_c\}} \sum_{c=1}^k \sum_{n \in N_c} w_n || x_c - x_n ||^2
  \end{equation}
  To lessen the risk of finding local minima the $k$-means algorithm
  is run on $10$ different starting conditions and all but the best
  found centroid configuration are discarded. Further, the clustering
  is constrained so that for each country and synchronous zone a
  number of clusters proportional to its overall mean load is chosen.
\item A new bus $c$ is created at each centroid $x_{c}$ to represent
  the set of clustered nodes.
\item All generators, storage units and loads that were connected to
  the original buses in $N_c$ are then aggregated by technology type
  at the equivalent bus $c$. The maximum expansion potential of
  generators of the same technology type are added for the new
  aggregated generator, while the weather-dependent availability time
  series for renewable generators are averaged with a weighting.
\item Lines between the clusters are replaced by a single equivalent
  line with a length of 1.25 times the crow-flies-distance, whose
  capacity is given by the sum of the capacity of the replaced lines,
  and whose impedance is given by the equivalent impedance of the
  parallel lines.
\end{enumerate}

Note that by focussing on the geographical distribution of the load
and conventional generation, this method ignores both the electrical
distance between the buses and the grid topology. Electrical distance
is ignored because the network clustering should be independent of
existing grid capacities, given that these capacities will be
optimised later; for the optimisation, geographical distance is more
important because it determines the cost of the grid expansion. The
topology is ignored because it is expected that the grid topology was
designed to connect major load and conventional generation centres, so
that focussing on the load and generation is sufficient to capture the
important conglomerations and the transmission corridors between them.

N-1 security is modelled by scaling down the available transmission
capacity to 0.7 times nominal capacity for high resolution networks
$\geq 200$ clusters and linearly shrinking it down until reaching half
the nominal values at $37$ clusters to account for cluster-internal
bottlenecks.

The network model is clustered down to
$k = 37, 45, 64, 90, 128, 181, 256, 362$ buses (see
Figure~\ref{fig:noexpansion-map} for the clusterings with $64$ and
$362$ buses) and the different results of the system optimisation are
examined for each level of clustering in several grid expansion
scenarios.

The network reduction algorithms are implemented in the free software
`Python for Power System Analysis (PyPSA)' Version 0.8.0 \cite{PyPSA},
which is developed at the Frankfurt Institute for Advanced Studies
(FIAS). PyPSA uses the scikit-learn Python package \cite{scikit-learn}
for the $k$-means clustering.

\section{Model for investment optimisation}

The model minimises total annual system costs, which include the
variable and fixed costs of generation, storage and transmission,
given technical and physical constraints.

To obtain a representative selection of weather and demand conditions
while keeping computation times reasonable, the model is run over
every third hour of a full historical year of weather and demand data
assuming perfect foresight, with 2012 chosen as the representative
year. Each time point $t$ is weighted by $w_t=3$ in the objective
function and storage constraints, to account for the fact that it
represents three hours.

The optimisation minimises total annual system costs, with objective
function
\begin{equation}
  \min_{\substack{G_{n,s},F_\ell,\\ g_{n,s,t},f_{\ell,t}}} \left[ \sum_{n,s} c_{n,s} G_{n,s} + \sum_{\ell} c_{\ell} F_{\ell} + \sum_{n,s,t} w_t o_{n,s} g_{n,s,t} \right]
\end{equation}
consists of the capacities $G_{n,s}$ at each bus $n$ for generation
and storage technologies $s$ and their associated annualised fixed
costs $c_{n,s}$, the dispatch $g_{n,s,t}$ of the unit in time $t$ and
the associated variable costs $o_{n,s}$, and the line capacities
$F_\ell$ for each line $\ell$ (including both high voltage alternating
current (HVAC) and direct current (HVDC) lines) and their annualised
fixed costs $c_\ell$.

The dispatch of conventional generators $g_{n,s,t}$ is constrained by
their capacity $G_{n,s}$
\begin{equation}
  0 \leq g_{n,s,t} \leq G_{n,s} \hspace{1cm} \forall\, n,s,t
\end{equation}

The maximum producible power of renewable generators depends on the
weather conditions, which is expressed as an availability
$\bar{g}_{n,s,t}$ per unit of its capacity:
\begin{equation}
 0 \leq  g_{n,s,t} \leq \bar{g}_{n,s,t} G_{n,s} \hspace{1cm} \forall\, n,s,t
\end{equation}

The energy levels $e_{n,s,t}$ of all storage units have to be
consistent between all hours and are limited by the storage energy
capacity $E_{n,s}$
\begin{IEEEeqnarray}{rCl}
  e_{n,s,t} & = & \eta_0^{w_t} e_{n,s,t-1} -  \eta_{1} w_t\left[g_{n,s,t}\right]^-+ \eta_{2}^{-1}w_t \left[g_{n,s,t}\right]^+ \nonumber \\
  & & + w_tg_{n,s,t,\textrm{inflow}} - w_tg_{n,s,t,\textrm{spillage}} \nonumber \\
0 & \leq &  e_{n,s,t} \leq E_{n,s}   \hspace{1cm} \forall\, n,s,t
\end{IEEEeqnarray}
Positive and negative parts of a value are denoted as
$[\cdot]^+= \max(\cdot,0)$, $[\cdot]^{-} = -\min(\cdot,0)$.  The
storage units can have a standing loss $\eta_0$, a charging efficiency
$\eta_1$, a discharging efficiency $\eta_2$, inflow (e.g. river inflow
in a reservoir) and spillage. The energy level is assumed to be
cyclic, i.e. $e_{n,s,t=0} = e_{n,s,t=T}$.

CO${}_2$ emissions are limited by a cap $\textrm{CAP}_{CO2}$, implemented using the
specific emissions $e_{s}$ in CO${}_2$-tonne-per-MWh of the fuel $s$
and the efficiency $\eta_{n,s}$ of the generator:
\begin{equation}
  \sum_{n,s,t} \frac{1}{\eta_{n,s}} w_t g_{n,s,t}\cdot e_{s} \leq  \textrm{CAP}_{CO2} \quad \leftrightarrow \quad \mu_{CO2} \label{eq:co2cap}
\end{equation}
In all simulations this cap was set at a reduction of 95\% of the
electricity sector emissions from 1990.

The (inelastic) electricity demand $d_{n,t}$ at each bus $n$ must be
met at each time $t$ by either local generators and storage or by the
flow $f_{\ell,t}$ from a transmission line $\ell$
\begin{equation}
  \sum_{s} g_{n,s,t} - d_{n,t} = \sum_{\ell} K_{n\ell} f_{\ell,t} \hspace{1cm} \forall\, n,t
\end{equation}
where $K_{n\ell}$ is the incidence matrix of the network. This
equation is essentially Kirchhoff's Current Law (KCL).

In this paper it is assumed that the linear load flow is a good
approximation for a well-compensated transmission network
\cite{Stot09,Brown}. To guarantee the physicality of the network
flows, in addition to KCL, Kirchhoff's Voltage Law (KVL) must be
enforced in each connected network.  KVL states that the voltage
differences around any closed cycle in the network must sum to
zero. If each independent cycle $c$ is expressed as a directed
combination of lines $\ell$ by a matrix $C_{\ell c}$ then KVL becomes
the constraint
\begin{equation}
  \sum_{\ell} C_{\ell c} x_{\ell} f_{\ell,t} = 0 \quad  \hspace{1cm} \forall c,t \label{eq:kvl}
\end{equation}
where $x_\ell$ is the series inductive reactance of line $\ell$. Note
that point-to-point HVDC lines have no cycles, so there is no
constraint on their flow beyond KCL.

The flows are also constrained by the line capacities $F_{\ell}$
\begin{equation}
  |f_{\ell,t}| \leq F_{\ell} \hspace{1cm} \forall\,\ell,t
\end{equation}
Although the capacities $F_{\ell}$ are subject to optimisation, no new
grid topologies are considered.

Since line capacities $F_{\ell}$ can be continuously expanded to
represent the addition of new circuits, the impedances $x_\ell$ of the
lines would also decrease. In principle this would introduce a
bilinear coupling in equation (\ref{eq:kvl}) between the $x_\ell$ and
the $f_{\ell,t}$. To keep the optimisation problem linear and
therefore computationally fast, $x_\ell$ is left fixed in each
optimisation problem, updated and then the optimisation problem is
rerun in up to 4 iterations to ensure convergence, following the
methodology of \cite{Hagspiel}.

In order to investigate the interactions of spatial scale with
transmission expansion, the sum of all transmission line capacities
(HVAC and HVDC) multiplied by their lengths $l_\ell$ is restricted by
a line volume cap $\textrm{CAP}_{\textrm{trans}}$, which is then
varied in different simulations:
\begin{equation}
  \sum_{\ell} l_\ell \cdot F_{\ell} \leq  \textrm{CAP}_{\textrm{trans}} \quad \leftrightarrow \quad \mu_{\textrm{trans}} \label{eq:lvcap}
\end{equation}
The caps are defined in relation to today's line capacities
$F_{\ell}^{\textrm{today}}$, i.e.
\begin{equation}
  \textrm{CAP}_{\textrm{trans}} = x \cdot \textrm{CAP}_{\textrm{trans}}^{\textrm{today}} = x \cdot \sum_{\ell} l_\ell \cdot F_{\ell}^{\textrm{today}}~.
\end{equation}

The discussion in Section~\ref{sec:results} starts off with the no
expansion scenario,
$\textrm{CAP}_{\textrm{trans}} =
\textrm{CAP}_{\textrm{trans}}^{\textrm{today}}$ so that no network
expansion is possible beyond today's line capacities. In this scenario
transmission bottlenecks restrict the exploitation of the best
renewable sites and the smoothing effects across the continent;
generation is forced to be more localised and renewable variability
may have to be balanced by storage. Then, five expansion scenarios are
studied by gradually easing the cap
$\textrm{CAP}_{\textrm{trans}} = x \cdot
\textrm{CAP}_{\textrm{trans}}^{\textrm{today}}$ with
$x = 1.125, 1.25, 1.5, 2, 3$ until reaching three times today's
transmission volume, which is already above the optimal value for
overhead lines at high numbers of clusters, as we will discuss in
Section~\ref{sec:cap}.

The optimisation model was also implemented in PyPSA.

\section{Data inputs}
\label{ref:datainputs}

\ra{1.05}
\begin{table}
\caption{Investment costs} \label{tab:costs}
\centering
\begin{tabular}{@{}lrlrr@{}}
\toprule
Quantity                &Overnight   &Unit & FOM & Lifetime \\
& Cost [\euro] & & [\%/a] & [a]\\
\midrule
Wind onshore    &1182   &kW\el &3 & 20   \\
Wind offshore  &2506   &kW\el  &3& 20  \\
Solar PV            &600   &kW\el &4 & 20   \\
Gas             &400    &kW\el  &4& 30  \\
Battery storage         &1275   &kW\el  & 3 & 20 \\
Hydrogen storage        &2070   &kW\el  & 1.7 &20 \\
Transmission line       &400    &MWkm & 2 & 40\\
\bottomrule
\end{tabular}
\end{table}

The network reduction and subsequent investment optimisation were run
on a full model of the European electricity transmission system.

The existing network capacities and topology for the ENTSO-E area
(including continental Europe, Scandinavia, the Baltic countries,
Great Britain and Ireland) were taken from the GridKit extraction
\cite{wiegmans_2016_55853} of the online ENTSO-E Interactive
Transmission Map \cite{interactive}. The model includes all
transmission lines with voltages above 220~kV and all HVDC lines in
the ENTSO-E area (see Figure \ref{fig:europe-map}). In total the model
contains $5586$ HVAC lines with a volume of $241.3$~TWkm (of which
$11.4$~TWkm are still under construction), $26$ HVDC lines with a
volume of $3.4$~TWkm (of which $0.5$~TWkm are still under
construction) and $4653$ substations.

The hourly electricity demand profiles for each country in 2012 are
taken from the European Network of Transmission System Operators for
Electricity (ENTSO-E) website \cite{entsoe_load}. The geographical
distribution of load in each country is based on GDP and population
statistics for the NUTS3 regions.

Electricity generation in the model is allowed from the following
technologies: hydroelectricity, natural gas, solar PV, onshore wind
and offshore wind. Gas, solar and wind capacities may be expanded
within the model constraints.

Existing hydroelectricity capacities (including run-of-river,
reservoirs and pumped storage) were compiled by matching databases
CARMA~\cite{ummel2012-carmav3}, GEO~\cite{GEO}, DOE
Global Energy Storage Database~\cite{doestoragedb} and the PowerWatch
project coordinated by the World Resources
Institute~\cite{wripowerwatch}; no expansion of existing hydro
capacities is considered in the model.  The hydro energy storage
capacities are based on country-aggregated data reported by
\cite{kies2016,pfluger2011} and the inflow time series are provided by
\cite{kies2016}.

The only fossil fuel generators in the model are open cycle gas
turbines, whose efficiency is assumed to be 39\%. Their usage is
limited by the \co{} cap in equation (\ref{eq:co2cap}).

The potential generation time series for wind and solar generators are
computed with the Aarhus renewable energy atlas \cite{REAtlas} from
hourly historical weather data from 2012 with a spatial resolution of
$40 \times 40 \textrm{km}^{2}$ provided by the US National Oceanic and
Atmospheric Administration \cite{saha}.

The distribution of these generators is proportional to the quality of
each site given by the local capacity factor multiplied with the
maximum installable capacity of the site.  However, protected sites as
listed in Natura2000 \cite{natura2000} are excluded, as well as areas
with certain land use types from the Corine Land Cover database
\cite{corine2012}, as specified by \cite{Scholz}, to avoid building,
e.g., wind turbines in urban areas. The maximum water depths for
offshore wind turbines is assumed to be $50$~m.
The maximum installable capacity per bus and generator type is then
determined by scaling these layouts until the first site reaches a
maximum installation density of $2$ and
$1.7~\mathrm{MW}/\mathrm{km}^{2}$ for wind and solar,
respectively. These maximum densities are chosen conservatively to
take account of competing land use and minimum-distance regulations
for onshore wind turbines.


The model contains two extendable types of storage units: batteries
and hydrogen storage. Their charging and discharging efficiencies, as
well as cost assumptions for their power and energy storage capacities
are taken from \cite{budischak2013}.  It is assumed that the charging
and discharging power capacities of a unit are equal, and the energy
capacity $E_{n,s}=h_{max,s}*G_{n,s}$ is proportional to this power
capacity. The factor $h_{max,s}$ determines the time for charging or
discharging the storage completely at maximum power, and is set to
$h_{max}=6\,\mathrm{h}$ for batteries and to $168\,\mathrm{h}$ for H2
storage.

Investment and fixed operation and maintenance (FOM) costs for all
assets are listed in Table \ref{tab:costs}.  The costs for generating
assets are based on predictions for 2030 from DIW
\cite{schroeder2013}; the costs for battery and hydrogen electricity
storage power capacity
and energy storage capacity come from \cite{budischak2013}.  Although
the costs of lines $c_\ell$ are set to zero, as they are dual to the
line volume cap, these costs are added in afterwards in the results.
For the annualisation of overnight costs a discount rate of $7\%$ is
used. Gas variable costs add up to $21.6$ \euro/MWh\th
\cite{schroeder2013}.

\section{Results}
\label{sec:results}

\begin{figure}
  $\vcenter{\hbox{\includegraphics[width=8.9cm]{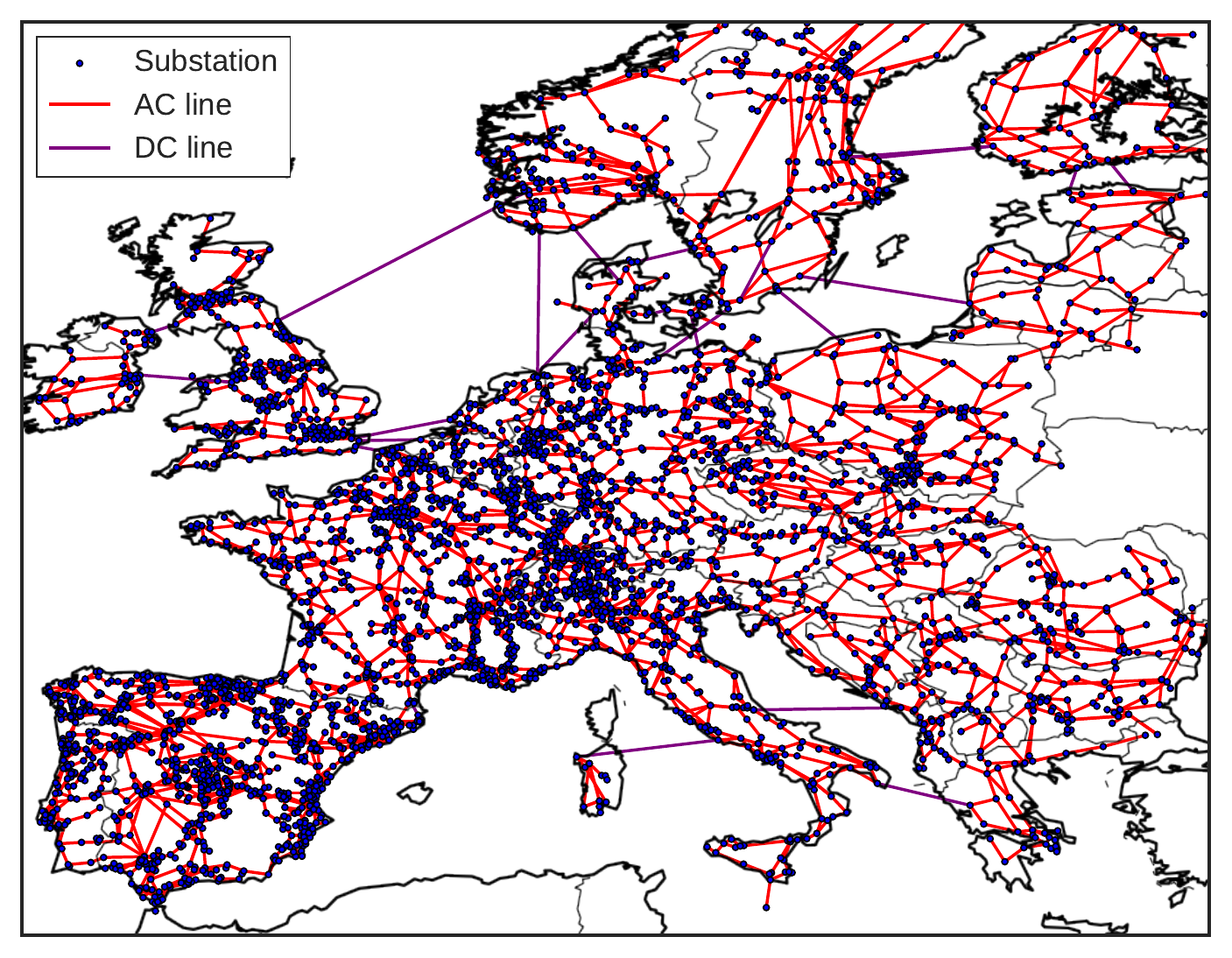}}}$
  \caption{Original grid model of Europe, described at the beginning of Sec.~\ref{ref:datainputs}}
  \label{fig:europe-map}
\end{figure}

\begin{figure}
  \includegraphics[width=8.9cm]{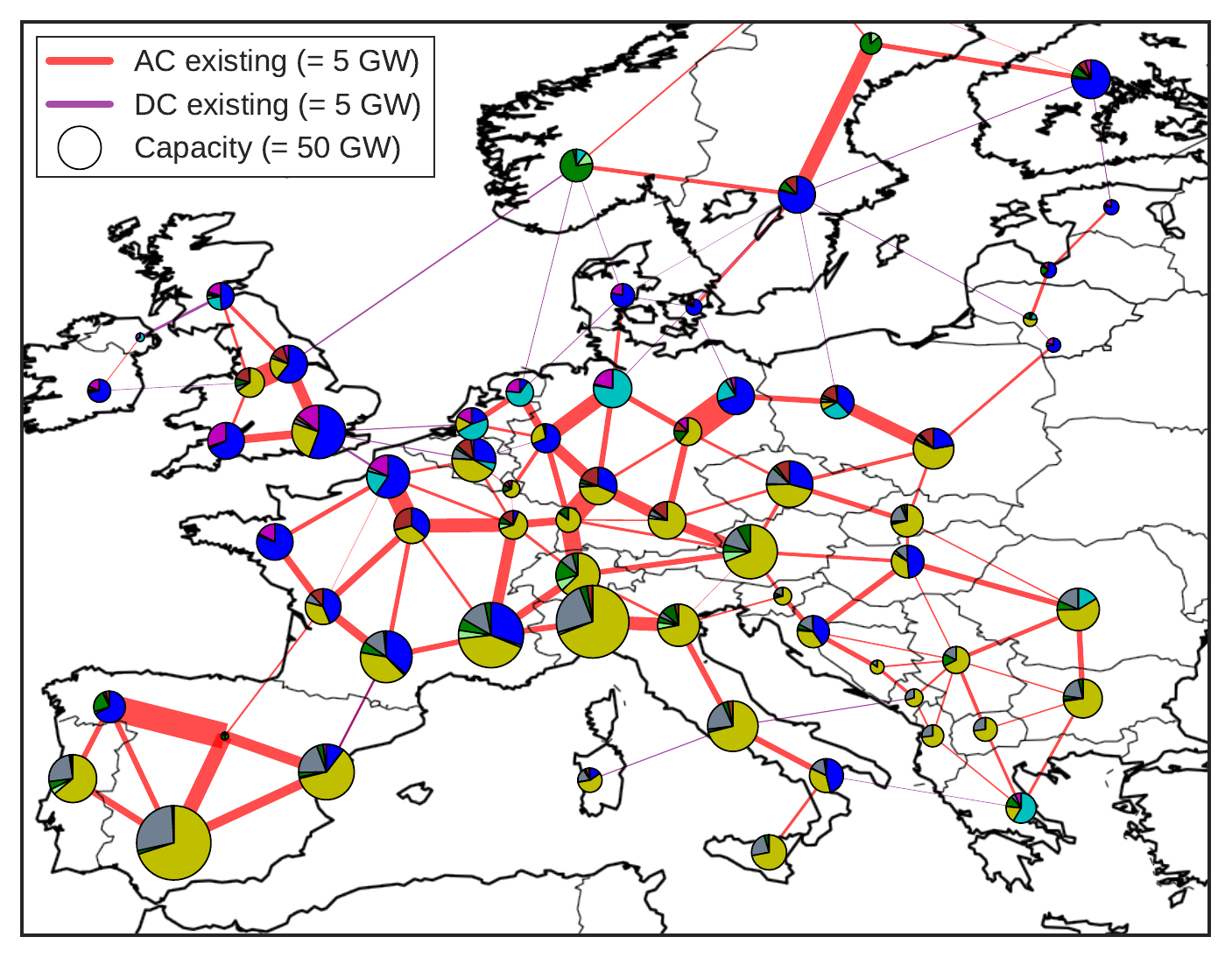}

  \includegraphics[width=8.9cm]{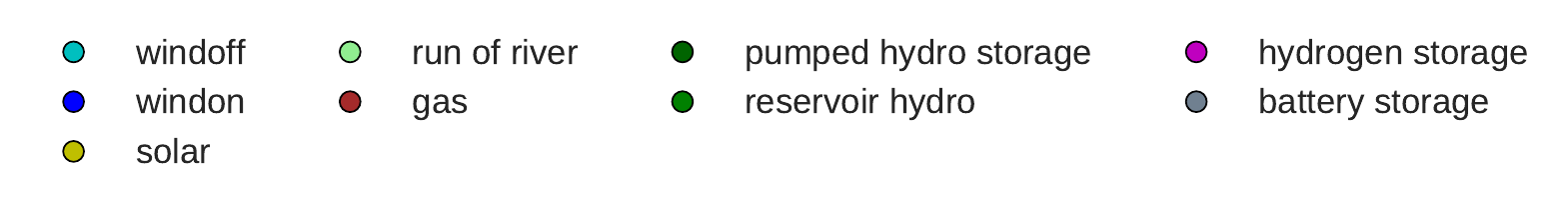} 

  \includegraphics[width=8.9cm]{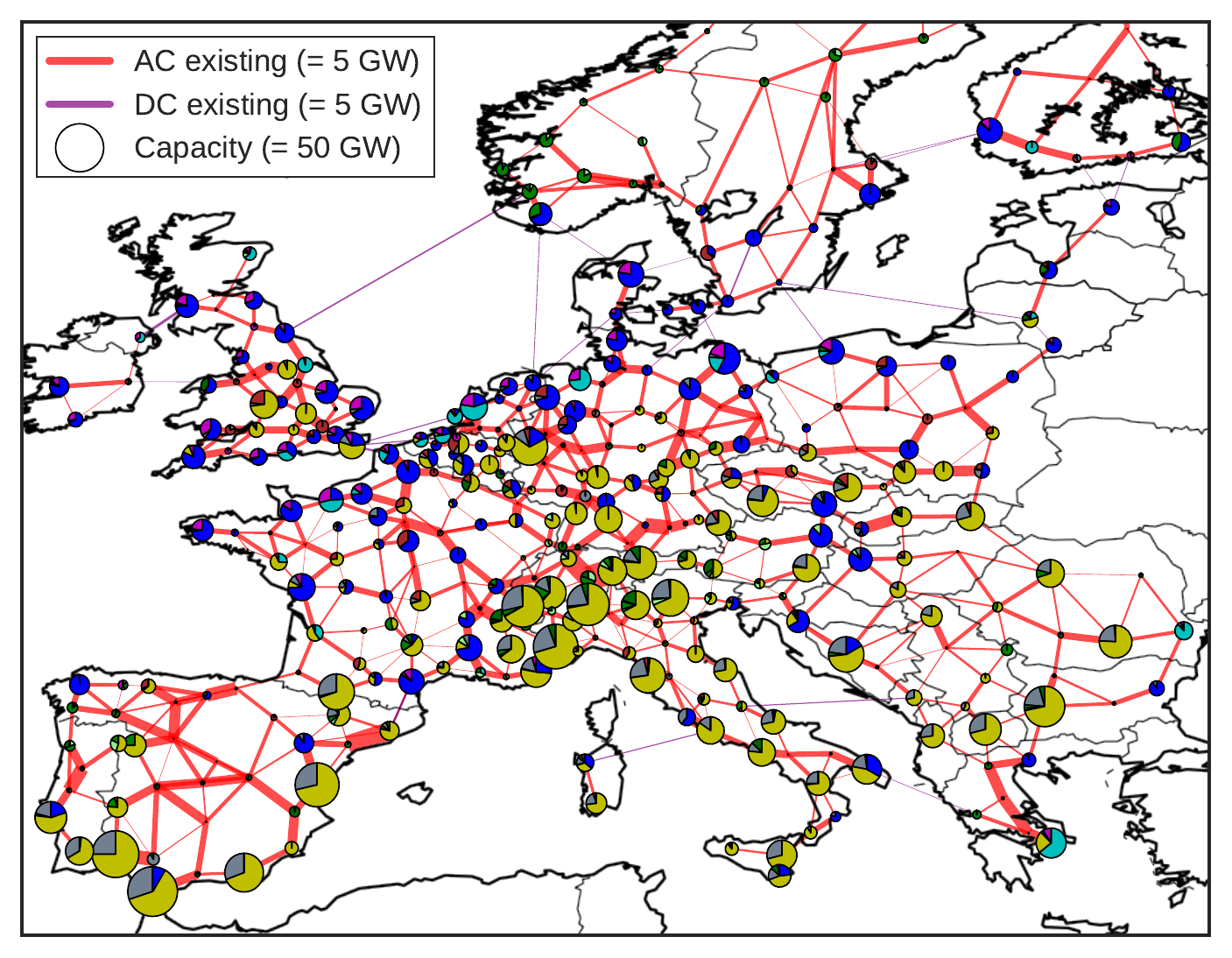}
  \caption{The clusterings with 64 buses (above) and 362 buses
    (below). Results for the distribution of generation capacities at
    each node are shown as pie charts for the no expansion scenario
    (existing and planned projects only).}
  \label{fig:noexpansion-map}
\end{figure}

\begin{figure*}
  \centering
  \includegraphics[scale=\figscale]{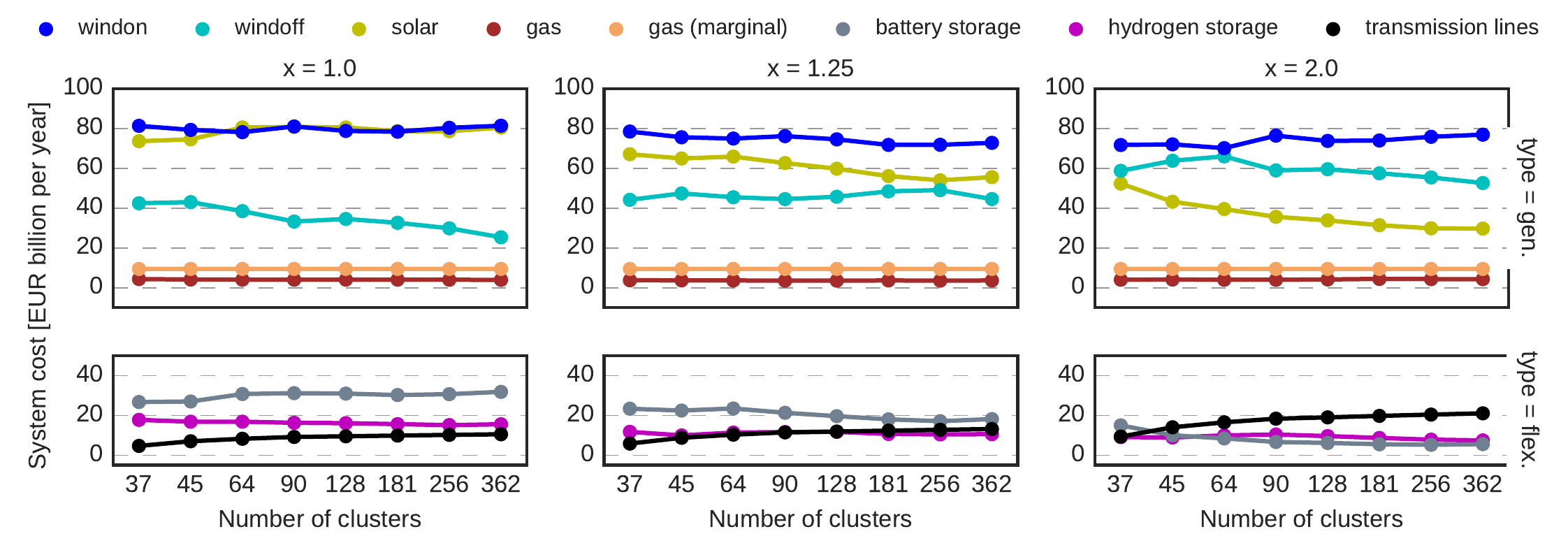}
  \caption{Breakdown of the annual system costs for generation (top) and flexibility options (bottom) as a function of the number of clusters for the no expansion scenario and the expansion scenarios with $x = 1.25$ and $x = 2$.}
\label{fig:costs-breakdown}
\end{figure*}

\begin{figure}[!t]
  \centering
  \includegraphics[scale=\figscale]{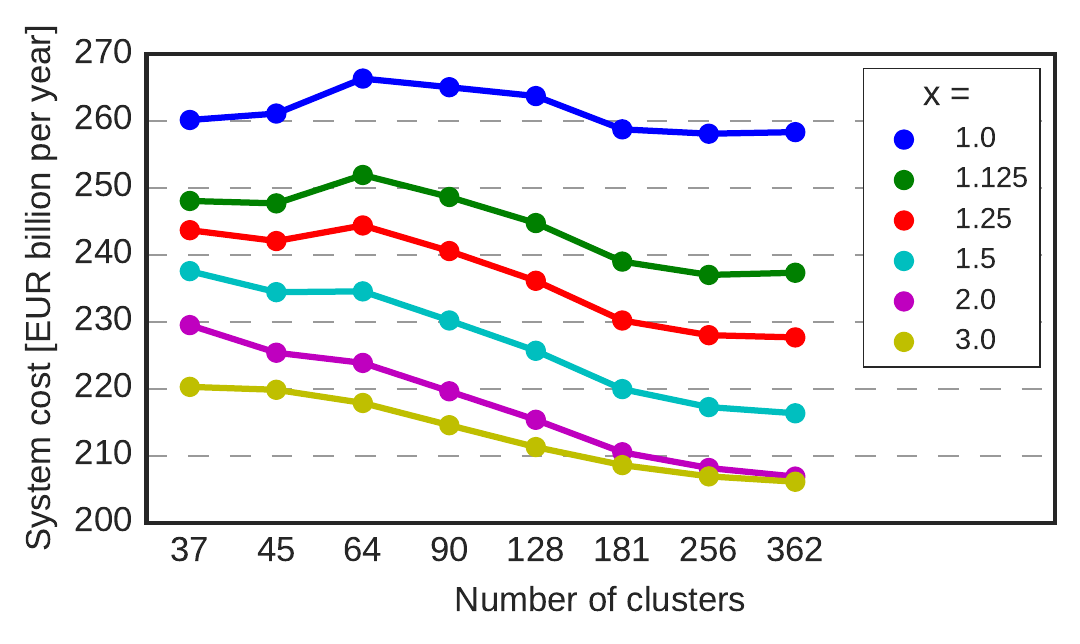}
  \caption{Total annual system costs as a function of the number of
    clusters for the six scenarios including the costs for overhead
    transmission lines.}
  \label{fig:costs}
\end{figure}

\begin{figure}
  \includegraphics[width=8.9cm]{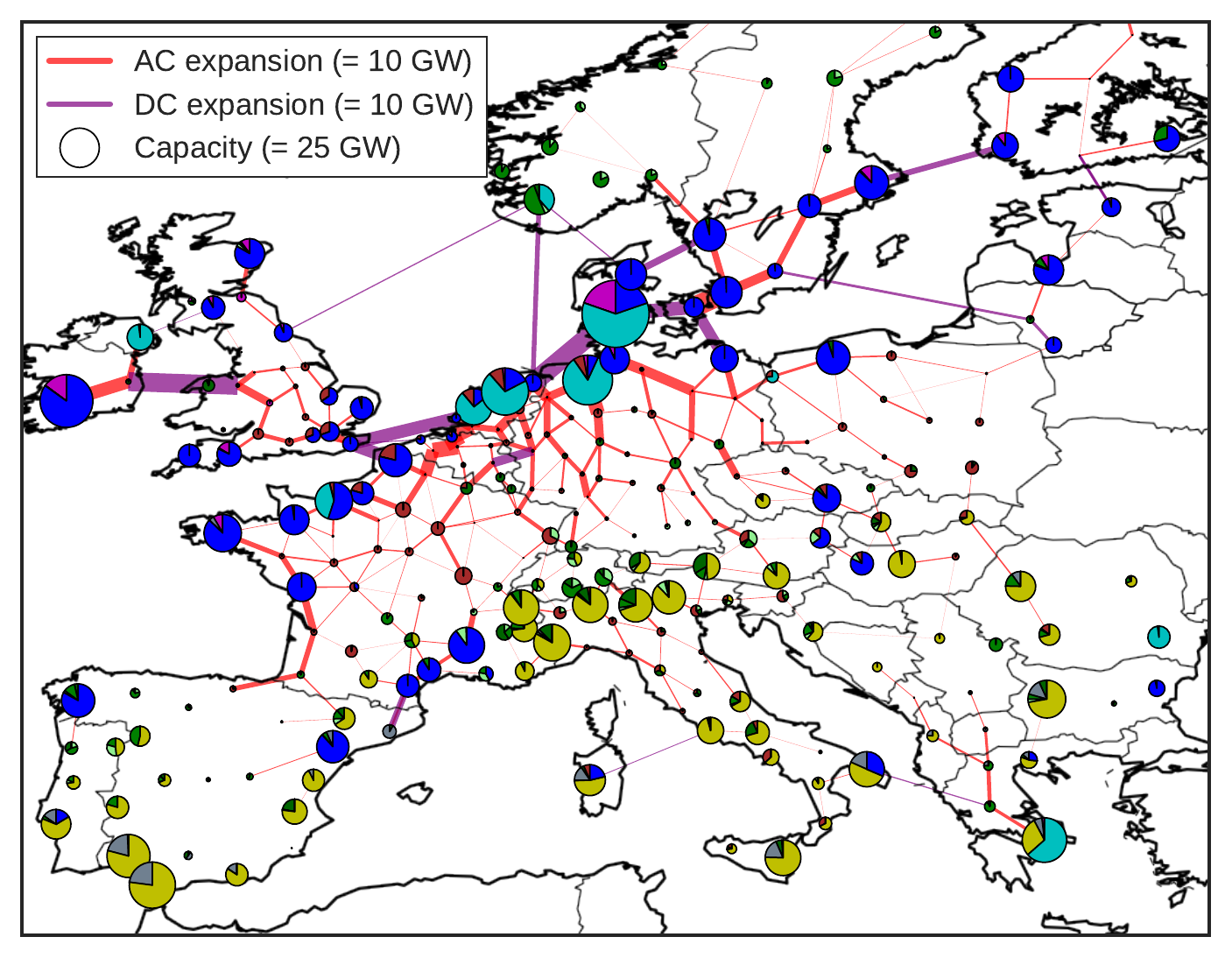}
  \includegraphics[width=8.9cm]{pre-7-legend}
  \caption{Optimal generation capacities and transmission line
    expansion for $256$ buses in the expansion scenario with
    the transmission cap at $x = 2$.}
  \label{fig:expansion-map}
\end{figure}

The original European grid model is shown in
Figure~\ref{fig:europe-map} and and can be compared to two clustered
networks in Figure~\ref{fig:noexpansion-map}; the total annual system
costs in the three scenarios as a function of the number of clusters
is found in Figure~\ref{fig:costs} and these costs are broken down
into components in Figure~\ref{fig:costs-breakdown}; the expansion of
the transmission network is shown in further detail in
Figure~\ref{fig:expansion-map}; the system costs as a function of the
transmission cap are plotted in Figure~\ref{fig:costs-per-cluster};
finally the shadow price of the transmission cap can be studied in
Figure~\ref{fig:shadows-branch_limit}. The results of the scenarios
are now discussed in detail.

\subsection{Spatial scale dependence}

Without any expansion of the transmission network ($x = 1.0$ in
Figure~\ref{fig:costs}), the total annual sytem cost remains
approximately steady as the number of clusters increases at 260
billion euros (an average of around \euro~82/MWh), due to a
coincidental balance of the two driving effects: (1) The sites with
high capacity factors are more finely resolved with a higher number of
clusters, allowing the model to put more capacity at the best
sites. With smaller numbers of clusters, the capacity factors are
averaged with a weighting over a larger area, bringing the capacity
factors down. For example, the best cluster for onshore wind in
Germany with 362 clusters has a capacity factor of about 40 \%,
whereas with one node for the whole of Germany, the weighted average
capacity factor is only 26 \%. (2) As the number of clusters increases,
the bottlenecks inside each country's network become constraining and
prevent the wind generated at high capacity factors, localised on the
coastlines and offshore, to be transported to load centres.

In the left panel in Figure~\ref{fig:costs-breakdown}, the two
effects, the increasing effective capacity factor of onshore wind
combined with intra-country bottlenecks becoming more important, lead
to the considerable decrease in the built offshore wind capacity,
since better sited onshore wind and solar installations produce more
energy closer to the load. The increasing solar generation drives an
increase in battery capacities to smooth short-term diurnal
variability. Hydrogen storage, which balances longer-term synoptic and
seasonal variability, decreases gently with the number of clusters at
a higher level than the other two scenarios. Gas generation is fixed
because of the \co constraint. The grid costs increase monotonically
as more line capacity and line constraints are seen by the model, but
flatten out with the exponentially increasing number of clusters. This
is a good indication that the clustering is capturing the major
transmission corridors even with smaller numbers of clusters.

Turning back to Figure~\ref{fig:costs}, the expansion of the network
lifts transmission bottlenecks and the first effect wins out, better
exploitation of good sites with higher numbers of clusters decrease
the system cost. As the grid is gradually expanded the system cost
decreases in a very non-linear manner: The expansion by $25\%$ reduces
the total system cost already by 30 billion euros of the 50 billion
euros in cost reduction available down to 210 billion euros (an
average of \euro~66/MWh). Nevertheless the overall cost reduction
possible by expanding the network is a moderate $20\%$.

In the technology break-down in the center and right panels of
Figure~\ref{fig:costs-breakdown}, with the additional line volume the
joint solar and battery capacities are replaced by offshore wind
turbines. Solar is favoured with limited transmission capacity because
it can be built close to demand everywhere and reasonably balanced
during its principal short-term diurnal variation using battery
storage, whereas the good wind sites are concentrated in Northern
Europe and their energy cannot be transported to loads in large
quantities without an expansion of the transmission grid. Wind
generation additionally benefits from expanding the transmission
capacities so that the spatial variation on the continental scale is
used for smoothing the temporal fluctuations on the synoptic scale to
relieve expensive hydrogen storage. The extra transmission capacity
does not offset the low significance of the transmission network cost.

These trends are all pronounced if the results for Germany are
considered in isolation. Transmission bottlenecks within Germany
complicate transporting offshore wind energy away from the coast with
higher numbers of clusters, forcing a dramatic substitution by solar
instead, i.e. the German offshore wind capacity falls from $40$ GW to
$12$ GW from 37 to 362 clusters, while solar peak capacities increase
from $46$ GW to $100$ GW and onshore wind remains largely unaffected
despite an intermediate decrease.

The effects disappear for about 200 clusters and above, a level of
resolution above which all the results are more-or-less steady.

\subsection{Transmission volume cap}
\label{sec:cap}

After ensuring that the solutions have already stabilized at 200
clusters and are thus, likely, a good proxy for the relations on the
full network, we want to focus in more detail on the solutions for 256
clusters while varying the allowed overall transmission volume to find
the most important lines for expansion and estimate the benefits of a
partial expansion deviating from the optimal solution, which might be
preferable vis-à-vis problems of public acceptance.

\begin{figure}
  \centering
  \includegraphics[scale=\figscale,trim=0 0.3cm 0 0]{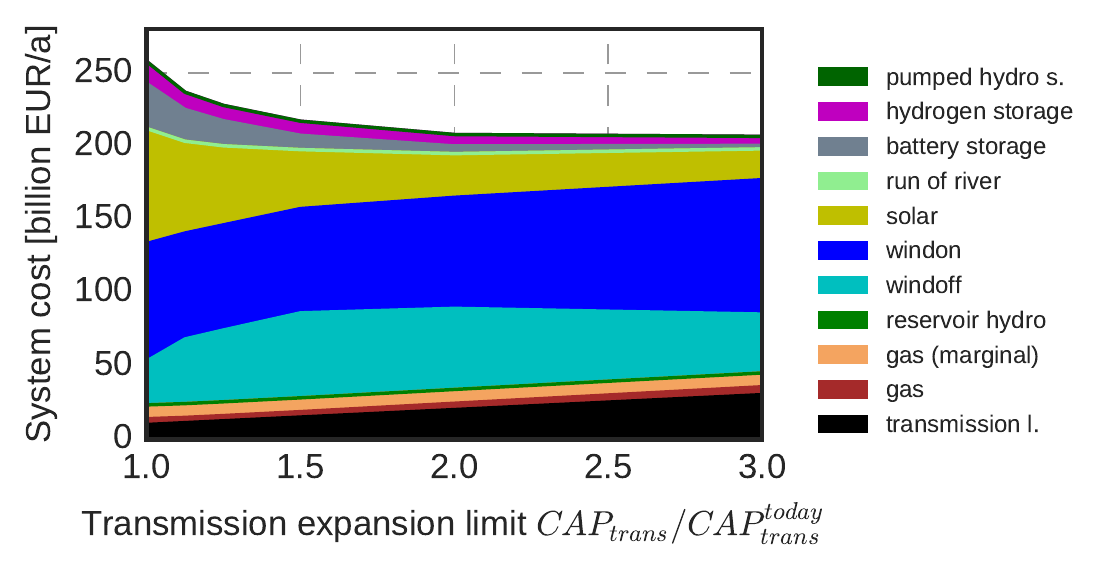}
  \caption{Annual total system cost at 256 clusters for different
    values of the transmission cap $\text{CAP}_{\text{trans}}$.}
  \label{fig:costs-per-cluster}
\end{figure}

\begin{figure}
  \centering
  \includegraphics[scale=\figscale,trim=0 0.3cm 0 0]{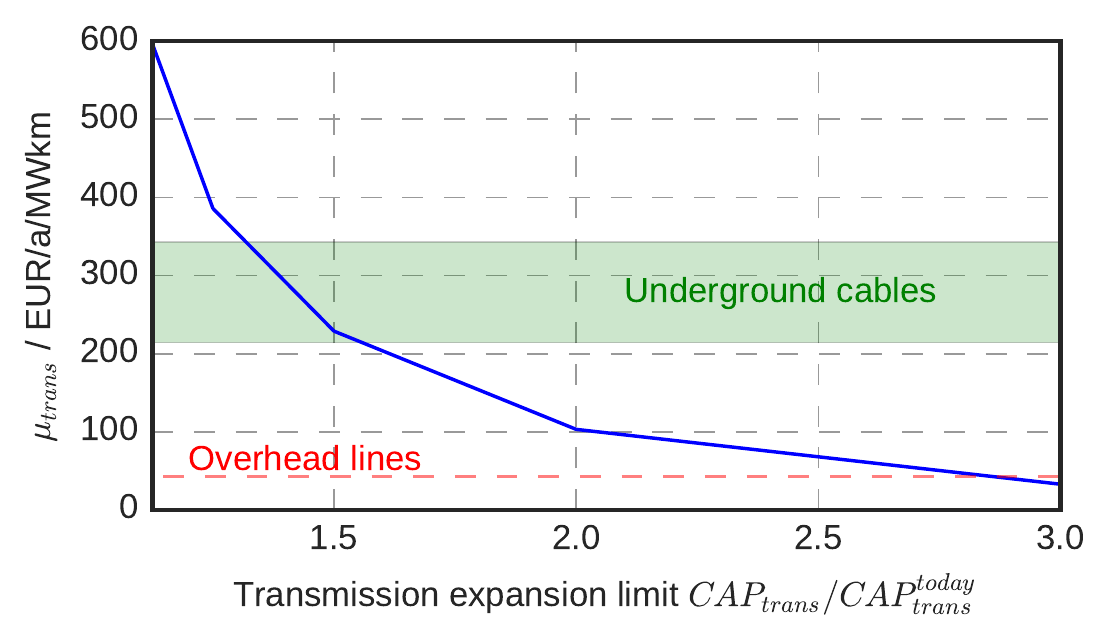}
  \caption{Shadow price of the line volume constraint
    $\mu_{\text{trans}}$ for different values of the transmission cap
    $\text{CAP}_{\text{trans}}$ for 256 clusters.}
  \label{fig:shadows-branch_limit}
\end{figure}

Figure~\ref{fig:expansion-map} shows the optimal generation capacities
and transmission expansion for a challenging doubling of the existing
transmission capacity ($x = 2$), which was also the subject of the
right panel in Figure~\ref{fig:costs-breakdown}. Transmission is
foremost expanded in the proximity of wind capacity installations
forming a wide band along the shore of the north and east sea with
branches leading inland. This band allows synoptic-scale balancing as
weather systems pass from west to east over Europe. It provides the
flexibility for the energy of large-scale wind installations to
replace a significant amount of solar capacity in Southern Europe and
Italy in particular, which also lessens the need for short-term
battery storage.

The total system cost in respect to allowed transmission volume in
Figure~\ref{fig:costs-per-cluster} decreases non-linearly as has
already been observed in the detailed study in the
one-node-per-country setting by
Schlachtberger~et.~al~\cite{schlachtberger2017}. More than half of the
overall benefit of transmission of $50$ billion EUR per year is
already locked in at an expansion by a fourth to
$1.25 \cdot \text{CAP}_{\text{trans}}^{\text{today}}$ and after
reaching two times today's line volume ($x = 2$) does not increase
significantly anymore (also compare the vertical slice at 256 clusters
in Figure~\ref{fig:costs}). From a system constrained to today's
transmission capacities to the optimal solution, the cost composition
reduces the component spent on solar and battery in favour of offshore
wind and, then, also onshore wind.

Finally, the shadow price of the transmission cap, $\mu_{trans}$
introduced in Equation~\eqref{eq:lvcap} is shown in
Figure~\ref{fig:shadows-branch_limit}. It indicates the marginal value
of an increase in line volume at each level of network expansion; it
can also be interpreted as the transmission line cost per MWkm
necessary for the optimal solution to have the transmission line
volume $\text{CAP}_{\text{trans}}$. For the assumed costs for
over-land transmission lines of $400$ \euro/MWkm the model finds the
optimal grid volume at slightly below
$3\cdot\text{CAP}_{\text{trans}}^{\text{today}}$. If the expansion
instead were to be carried out with underground cabling at a 4 to 8
times higher cost, the economically optimal solution would still be to
expand the line volume to between $1.25$ and $1.5$ times the existing
volume.

\section{Critical appraisal}

Although the clustering algorithm presented in Section
\ref{sec:reduction} captures the major transmission corridors well, it
would be interesting to benchmark the different clustering algorithms
mentioned in the introduction based on comparable criteria, such as
their ability to capture power flows in the original unclustered
network. Results with a higher number of nodes would also be
desirable, if this is computationally possible.

Additional aspects, such as distribution grid costs, reserve power,
stability and sector-coupling, have not been considered here.

\section{Discussion and Conclusions}

The results of this paper are two fold: Firstly, a network clustering
method has been demonstrated that can reduce the number of buses in a
given electricity network while maintaining the major transmission
corridors for network analysis. With this network reduction method the
effects of spatial resolution, i.e. the number of clusters, on the
joint optimisation of transmission and generation investment for
highly renewable systems in Europe have been investigated. Secondly,
the techno-economic European model was optimised at a sufficient level
of resolution to determine the hotspots and benefits of transmission
expansion.

The systems optimised to reduce \co{} emissions by 95\% with no grid
expansion are consistently only around 20\% more expensive than
systems with grid expansion and half of that cost benefit can already
be locked in with an expansion of the line volume by a fourth, which
may be a price worth paying given public acceptance problems for new
transmission lines.

One must note, though, that in the time horizon until 2050 in which
the studied reduction of emissions is to be implemented a significant
amount of the current conventional generation park will not yet have
passed their lifetime and an important next step is to confirm our
greenfield results accounting for this inertia. Further, one should be
clear that the feasibility of these solutions is based on a fully
integrated European market with nodal prices, high \co{} price and
optimally real-time prices for distributed generation and storage.


\section*{Acknowledgments}

This research was conducted as part of the CoNDyNet project, which is
supported by the German Federal Ministry of Education and Research
under grant no. 03SF0472C. The responsibility for the contents lies
solely with the authors.



%

\printbibliography

\end{document}